\documentclass[%
 aapm,
 mph,%
 amsmath,amssymb,
 reprint,%
]{revtex4-2}
\usepackage{subcaption}
\usepackage{amsmath}
\usepackage{graphicx}
\usepackage{ragged2e}
\usepackage{graphicx}
\usepackage{float}
\usepackage{xcolor}
\usepackage{subcaption}
\usepackage{textcomp}
\usepackage{soul}

\usepackage{booktabs}

\usepackage{upgreek}

\usepackage[utf8]{inputenc}

\usepackage[normalem]{ulem}
\usepackage{graphicx}
\usepackage{float}

\usepackage{subcaption}
\usepackage{textcomp}
\usepackage{soul}
\usepackage{ragged2e}
\usepackage{amsmath}
\usepackage{dblfloatfix}

\usepackage{hyperref}

\newcommand{\ket}[1]{|{#1}\rangle}
\newcommand{\bra}[1]{\langle{#1}|}

\usepackage{lineno}
\usepackage{graphicx}
\usepackage{dcolumn}
\usepackage{bm}

\begin{document}

\preprint{AAPM/123-QED}

\title{EIT in V+ inverted \texorpdfstring{$\Xi$}{Xi} system using Rydberg state in thermal Rb atoms}

\author{Thilagaraj Ravi}
\author{Heramb Vivek Bhusane}%
\author{Rajnandan Choudhury Das}%
\author{Samir Khan}%
\author{
Kanhaiya Pandey
}%
 \email{kanhaiyapandey@iitg.ac.in}
\affiliation{ 
Department of Physics, Indian Institute of Technology Guwahati, Guwahati, Assam 781039, India
}%


\begin{abstract}
Rydberg excitation using blue and IR transition is an advantageous path for quantum computation in alkali elements. Aiming to stabilize the IR laser for quantum computation, we study electromagnetically induced transparency (EIT) spectrum using Rydberg state in V+inverted $\Xi$ system (${5S_{1/2}}$ $\rightarrow$ ${5P_{3/2}}$ and ${5S_{1/2}}$ $\rightarrow$ ${6P_{1/2}}$  $\rightarrow$ ${r=69D_{3/2}}$) in Rb vapour cell at room temperature. The probe laser absorption at 780 nm is monitored in the presence of the two control lasers at 421 nm and 1003 nm. In comparison to the previously studied inverted $\Xi$ system, this system has a good signal-to-noise ratio even at room temperature with similar linewidth (around $10$~MHz). We also observe Autler-Towns splitting of the EIT due to high power of probe and blue control lasers. For completeness and comparison, we also study the EIT in an inverted $\Xi$ system using $5S_{1/2}\rightarrow6P_{1/2}\rightarrow 69D_{3/2}$ transitions. 
\end{abstract}
\date{\today}
\maketitle

For atomic platform, Rydberg excitation is key for the quantum computation \cite{lukin1,lukin2} and simulation \cite{scholl2021,browaeys2024} as it enables the multi-qubit operation through Rydberg blockade. In Rb, one of the extensively used path for Rydberg excitation is ${5S_{1/2}}$ $\rightarrow$ ${5P_{3/2}}$ at 780 nm and ${5P_{3/2}}$ $\rightarrow$ ${r}$ (Rydberg states) at around 480 nm. However, the other scheme ${5S_{1/2}}$ $\rightarrow$ ${6P_{3/2}}$ at 420 nm and ${6P_{3/2}}$ $\rightarrow$ ${r}$ at 1013 nm for Rydberg excitation is better for quantum computation. This is because (i) the narrow linewidth \cite{Das_2023,ryd420} of the intermediate state (${6P_{3/2}}$) reduces the decoherence rate for two-photon coherent Rydberg excitation, (ii) it is easy to produce high power 1013 nm laser.
 
The coherent Rydberg excitation requires frequency stability of Rydberg lasers and spectroscopic tool will be a good option for it. The stabilization of the blue laser is possible through saturated absorption spectroscopy (SAS) \cite{Nyakang2020, Ogaro21, abs_420,das_direct_2024} however the spectroscopy of IR laser is not possible through SAS as there is no population in the excited state ($6P$) and hence EIT spectrum will be a good option for it. 

EIT in thermal vapour for $^{87}$Rb atoms in $5S_{1/2}\rightarrow6P_{3/2}\rightarrow r$ (Rydberg state) inverted $\Xi$ system has been studied before \cite{Brekke:23}. In the inverted $\Xi$ system probe wavelength ($\lambda_{p}$  $\approx $ 421 nm) is considerably shorter than the control laser wavelength ($\lambda_{c}$  $\approx $ 1003 nm) and termed as inverted $\Xi$ system \cite{PhysRevA.101.052507}. Note that the systems with $\lambda_p > \lambda_c$ is termed as $\Xi$ system \cite{stl2021,stl2013,Moon2012}. The signal-to-noise ratio of the EIT signal for the inverted $\Xi$ system happens to be poor because (i) absorption of blue probe is weak, (ii) in the weak (first order) probe limit \cite{PAN13} there is no EIT at room temperature, however at high power (in higher order of probe) EIT exist \cite{Urvoy_2013,PhysRevA.101.052507,PhysRevA.105.042808}. For significant absorption of blue probe \cite{das_direct_2024} to detect a change caused by EIT, we need to heat the cell up to 80-90 $^o$C . Further, we also need a blue-sensitive photodetector.

To improve the signal-to-noise ratio, we investigate EIT spectrum in thermal vapor for $^{87}$Rb atoms in the V+inverted $\Xi$ system $5S_{1/2}\rightarrow5P_{3/2}+5S_{1/2}\rightarrow6P_{1/2}\rightarrow 69D_{3/2}$ with 780 nm probe and 421 nm and 1003 nm as control lasers. 
As the probe at 780 nm has stronger absorption as compared to the 421 nm probe, we get the EIT spectrum with a better signal-to-noise ratio even without heating the cell. Further, it also does not require a blue-sensitive photodetector. The probe ($5S_{1/2}, F = 2 \rightarrow5P_{3/2}, F=3$) absorption is monitored as scan of the $6P_{1/2}\rightarrow 69D_{3/2}$, 1003 nm control laser with fixed frequency of $5S_{1/2}\rightarrow6P_{1/2}$ \cite{Brekke:23}, 421 nm control laser. For completeness and comparison, we also study the inverted $\Xi$ system in this work. 

Further, the EIT has been theoretically studied in V+$\Xi$ system \cite{BHARTI2015510, kaur_study_2018} for various applications such as sub- and super-luminal light propagation \cite{BAN17}, all-optical grating \cite{PhysRevA.98.043822} but with limited experimental studies \cite{Yuan:20}. To best of our knowledge, this is the first experimental study on the V+ inverted \texorpdfstring{$\Xi$}{Xi} system.

\begin{figure}[ht!]
	\begin{subfigure}{\linewidth}
		\centering
        \includegraphics[width=0.9\linewidth]{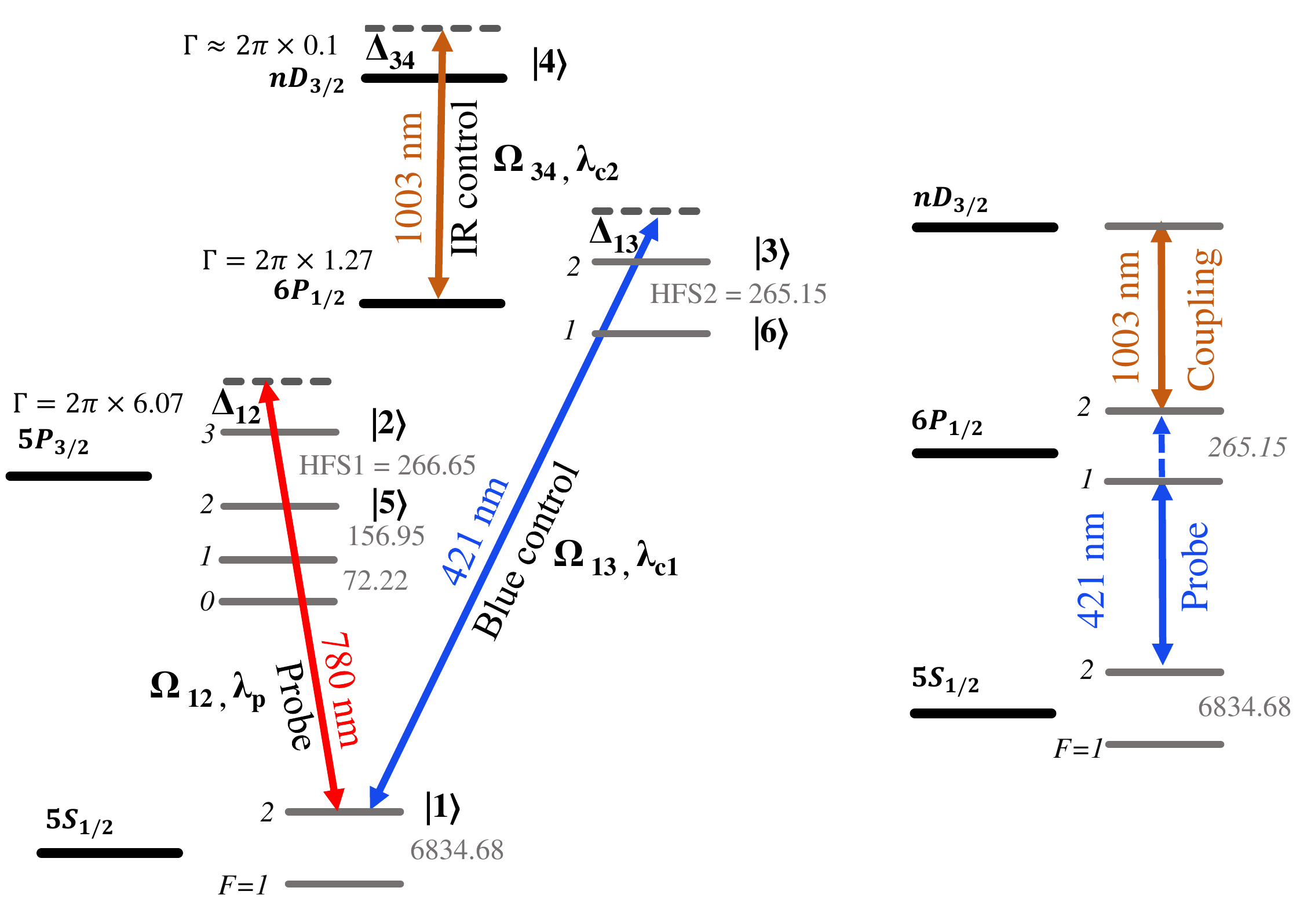}
	\end{subfigure}
	\begin{subfigure}[]{\linewidth}
		\centering
        \includegraphics[width=\linewidth]{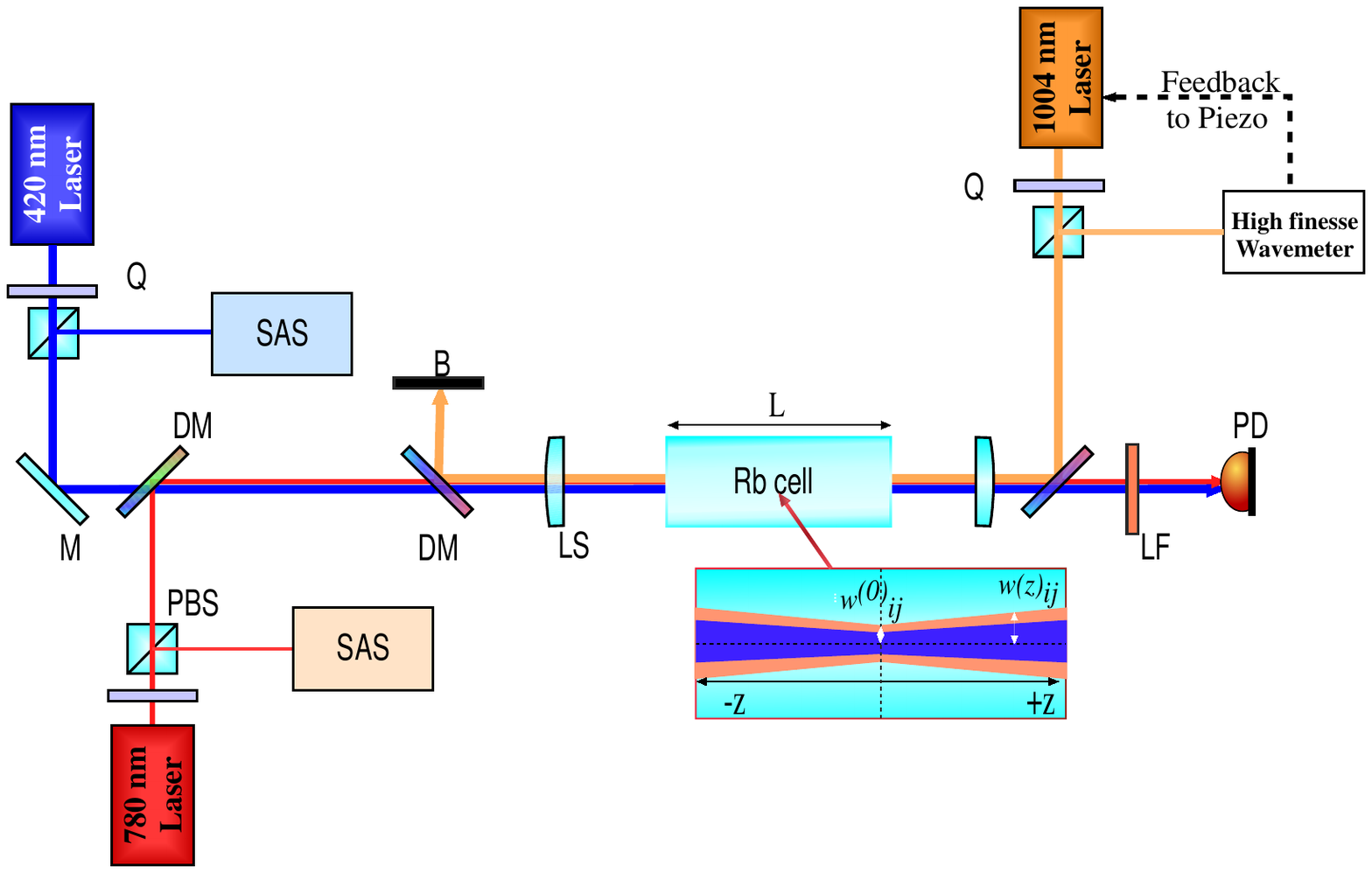}  %
			\label{fig:ladder_exp}
	\end{subfigure}
		\begin{picture}(0,0)
		\put(-100,330){(a)}
		\put(50,330){(b)}
		\put(-100,170){(c)}
	\end{picture}
	\caption{ \justifying \textbf{a (b).} Energy level diagram for V+inverted $\Xi$ system (inverted $\Xi$ system ) in $^{87}$Rb atoms. The numbers denoted are the hyperfine splitting of the levels in MHz. The value (in MHz)  of population decay rate of levels are mentioned above the levels \textbf{c.} Optics layout for the experimental setup. LS: Plano-convex lens, DM: Dichroic mirror, PD: Photodiode, Q: Half waveplate, B: Beam dump, M: Mirror, LF: Spectral filter}
	\label{fig:both}
\end{figure}

The energy level diagram for V+inverted $\Xi$ system is shown in Fig \ref{fig:both}a.
The experiment is conducted in a cylindrical quartz vapor cell of diameter 25~mm and length $L=5$~cm as shown in Fig \ref{fig:both}c. The blue control beam is derived from a home-built external cavity diode laser (ECDL) with a linewidth $<$ 500 kHz and output power of 70 mW, and the IR control beam is generated by another home-built ECDL of typical linewidth $<$ 500 kHz, with a laser diode (Model: SM-1000-TO-300, Make: Innolume) that can deliver up to 200 mW power. The probe, blue control, and IR control beams are focused at the center of the vapor cell with typical beam waists of $\omega^{(0)}_{12}=$ 70 $\mu$m, $\omega^{(0)}_{13}=$ 30 $\mu$m and $\omega^{(0)}_{34}=$ 70 $\mu$m, respectively, and spatially overlapped in the vapor cell using dichroic mirrors. The 780 nm probe is stabilized at $5S_{1/2}$ F=2 $\rightarrow$ $5P_{3/2}$ F=3 transition using SAS. The IR control addresses various Rydberg levels and its wavelength is monitored by a wavemeter (WS7-60, High Finesse) while SAS setup is used to stabilize the blue control $5S_{1/2}$ F=2 $\rightarrow$ $6P_{1/2}$ F=2 transition \cite{das_direct_2024}. The probe 780 nm and blue control beams are co-propagating and IR control beam is counter-propagating to both inside the cell.

The energy level for the inverted $\Xi$ system is shown in Fig \ref{fig:both}b. In this case blue probe laser is stabilized to $5S_{1/2} F=2 \rightarrow 6P_{1/2}$ $F = 2$ transition using double resonance spectroscopy \cite{Nyakang2020}. The IR coupling laser is scanning and addressing $6P_{1/2}$ $F = 1,2$ $\rightarrow 69D_{3/2}$. The blue probe transmission is monitored using a blue-sensitive photodetector. The vapour cell is heated to 80 \textdegree{}C. 

\begin{figure}[ht!]
	\centering

    \includegraphics[width=0.85\linewidth]{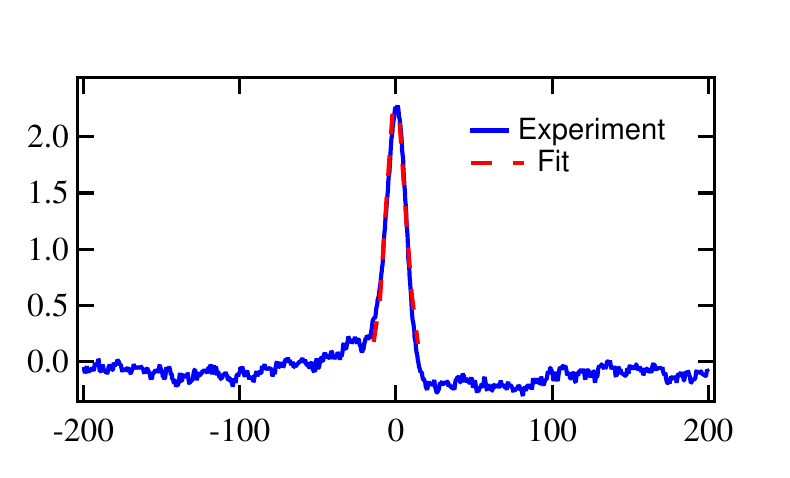}
	\begin{picture}(0,0)
		\put(-230,25){\rotatebox{90}{\shortstack{Probe transmission \\ (arb.units)}}}
		\put(-160,-1){IR laser detuning (MHz)}
	\end{picture}
\caption{ \justifying Probe transmission vs IR control laser detuning. 
Blue curve (experiment after averaging of eight traces) with the 780 nm probe, blue control, and IR control laser powers are 7 µW (calculated $\Omega^{(0)}_{12}$= 15 MHz), 0.2 mW (calculated $\Omega^{(0)}_{13}$= 8 MHz), and  40 mW (calculated $\Omega^{(0)}_{34}$=   16 MHz), respectively. Rabi frequencies ($\Omega^{(0)}_{ij}$) are calculated using ARC\cite{ARC} based on the cross-sectional average intensity for the minimum waist. Red curve (fitting): The extracted Rabi frequency from fitting $\Omega^{(0)}_{12}$ = 10 MHz, $\Omega^{(0)}_{13}$ = 11 MHz and $\Omega^{(0)}_{34}$=16 MHz.}
	\label{Vplusladder_spectrum}
\end{figure}
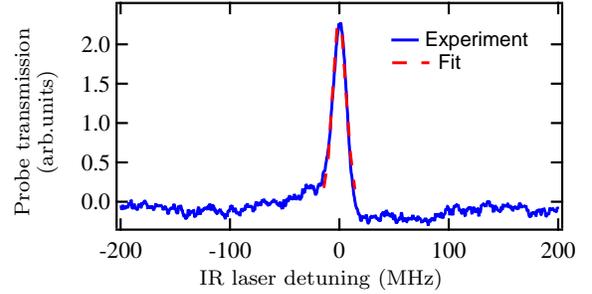

We now discuss the spectrum of the V+inverted \texorpdfstring{$\Xi$}{Xi} system. The transparency window (experimental and theoretical fitting) of 780 nm probe with scan of the IR control laser is shown in Fig. \ref{Vplusladder_spectrum}. For low power of 780 nm probe and blue control lasers, the spectrum can be understood by population transfer to long-lived $69D_{3/2}$ Rydberg state through step excitation by blue and IR control lasers. As the blue laser is fixed to $5S_{1/2}F=2 \rightarrow 6P_{1/2}F=2$ transition, we observe a transparency window with a scan of the IR control laser. At lower probe power, we observe only one peak (unlike inverted $\Xi$ system) even though IR control laser is scanning across both the hyperfine levels, $6P_{1/2}$, F=1 and F=2. This is because only zero velocity group of atoms can be in resonance with both 780 nm probe and blue control laser. Even though the hyperfine separation between $F=3$ and $F=2$ of $5P_{3/2}$ and $F=2$ and $F=1$ of $6P_{1/2}$ is similar but due to different wavelengths of the probe and blue control lasers there is no velocity group which can make the probe in resonance with $5S_{1/2}F=2\rightarrow5P_{3/2}F=2$ and also blue control  $5S_{1/2}F=2\rightarrow6P_{1/2}F=1$. The signal to noise ratio ($(V_\textrm{signal})^2/(V_\textrm{noise})^2$) for the spectrum is 64 with eight averaging and for the error signal 25 without averaging.

The V+inverted $\Xi$ system can be modelled as four-level system $\ket{1}\rightarrow\ket{2}$ and $\ket{1}\rightarrow\ket{3}\rightarrow\ket{4}$ shown in Fig  \ref{fig:both}a. This system can be theoretically studied using density matrix analysis to compare it with experimental spectral features. The Hamiltonian for the system can be written as

\begin{align}
	\label{eq1}
	H&=\hbar[-\delta_{12}\ket{2}\bra{2}-\delta_{13}\ket{3}\bra{3}-(\delta_{13}+\delta_{34})\ket{4}\bra{4}\\ \nonumber
	&+ \frac{\Omega_{12}}{2}\ket{1}\bra{2}+\frac{\Omega_{13}}{2}\ket{1}\bra{3}+\frac{\Omega_{34}}{2}\ket{3}\bra{4}+h.c.]
\end{align}  

where the $\delta_{12}=\Delta_{12}+k_{12}v$, $\delta_{13}=\Delta_{13}+k_{13}v$, and $\delta_{34}=\Delta_{34}-k_{34}v$ for the atoms moving with $v$ velocity along the direction of the propagation of IR control laser. Here, $k_{12}=1/780~\textrm{nm}$, $k_{13}=1/421~\textrm{nm}$ and $k_{34}=1/1003~\textrm{nm}$ and $\Delta_{12}$, $\Delta_{13}$, and $\Delta_{34}$ are detuning of the 780 nm, 421 nm and 1003 nm lasers for stationary atoms. $\Omega_{12}$, $\Omega_{13}$, and $\Omega_{34}$ are the Rabi frequencies.

The dynamics of the system can be studied using  the Liouville- Von Neumann equation,
\begin{align}
	\frac{d\rho}{dt}=-\frac{i}{\hbar}[H,\rho]+L(\rho)
\end{align}
 
The above equation contains 16 coupled differential equations. We solve the above equation numerically with initial condition $\rho_{ij}=0$ $\forall$ $i$ and $j \in \{1,2,3,4\}$ except $\rho_{11}=1$, to find out the $\rho_{21}$ whose imaginary part is related to the absorption of the 780 nm probe laser. We do averaging for the longitudinal and transverse velocities w.r.t laser propagation direction. The longitudinal velocity effects $\rho_{21}$ through Doppler effect and transverse velocity causes finite interaction time given as $2\omega_{ij}(z)/v_t$, where $\omega_{ij}(z)$ is beam waist at position $z$ see Fig. \ref{fig:both}c. To save the computation, we take constant Rabi frequency in the cross-section of the laser beams and vary along the propagation direction using the Gaussian beam propagation formula. The averaged $\rho_{21}$ is given by  
\begin{align}
\label{ave}
\rho^{\textrm{ave}}_{21} &= A \int^{v_p}_{0}dv_t 
\int^{v_p}_{-v_p} dv_l \int_{\frac{-L}{2}}^{\frac{L}{2}} dz \\
\nonumber
&\quad \rho_{21}(v_t,v_l,\Omega_{12}(z), \Omega_{13}(z), \Omega_{34}(z)) \\
\nonumber
&\quad \times v_t e^{-\frac{m(v_t^2+v_l^2)}{2k_BT}} \pi \times (\omega_{ij}(z))^2
\end{align}

where, $v_p=\frac{2k_BT}{m}$, $A=\sqrt{1/2\pi}(m/k_B T)^{3/2}$ and $\Omega_{ij}(z)=\Omega^{(0)}_{ij}[1+(z/z_R)^2]$. $\Omega_{ij}(z)$ is Rabi frequency at position $z$ for $\ket{i}\rightarrow\ket{j}$ transition. $L$ is the length of Rb cell. The numerical integration is done with step size $\Delta v_t=\Delta v_l=2$m/s and $\Delta z=L/8$.  

We fit experimental data points (plotted with the blue curve) of Fig. \ref{Vplusladder_spectrum} using the above numerical solution with $\Omega^{(0)}_{12}$, $\Omega^{(0)}_{13}$ and $\Omega^{(0)}_{34}$ fitting parameters. To save computations, we use only 18 experimental data points in the fitting region from $-14$ MHz to $14$ MHz. The fitting is shown with red curve. There is mismatch between the extracted Rabi frequencies from fitting and the calculated values. This could be due to misalignment of the laser beams, the presence of multiple magnetic sub-levels, and the imperfect Gaussian beam shape of the lasers.

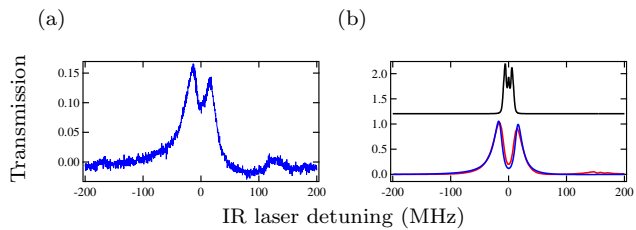
\begin{figure}[ht!]
	\centering
    \includegraphics[width=0.45\linewidth]{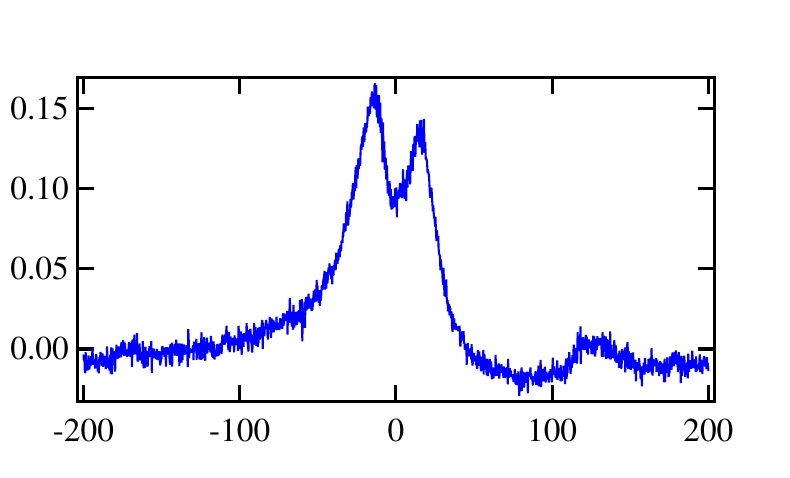}
    	\begin{picture}(0,0)
        \put(-120,70){(a)}
		\put(-130,10){\rotatebox{90}{\shortstack{Transmission}}}
		\put(-50,-5){IR laser detuning (MHz)}
	    \end{picture}
     \includegraphics[width=0.45\linewidth]{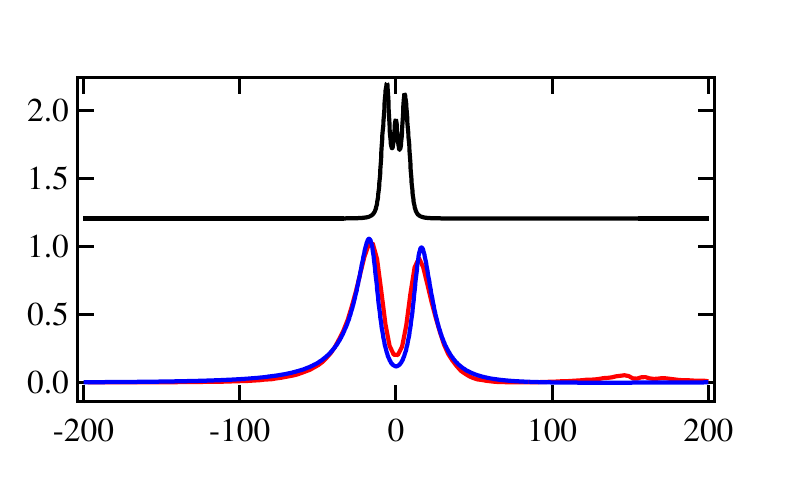}
   	\put(-102,40){\rotatebox{90}{}}
   \put(-95,-5){}
	  \put(-120,70){(b)}
 
	\caption{ \justifying 780 nm probe transmission (arb.units) vs IR control detuning for high probe power \textbf{(a)} Experiment:  The probe 780 nm probe, blue control, and IR control laser powers are  0.4 mW (calculated $\Omega^{(0)}_{12}$=  117 MHz), 7 mW (calculated $\Omega^{(0)}_{13}$=  45 MHz), and  40 mW (calculated $\Omega^{(0)}_{34}$ = 16 MHz), respectively. \textbf{(b)} Theory: The red (blue) solid line corresponds to theoretical prediction with Doppler averaging (stationary atoms) with $\Omega^{(0)}_{12}$=  50 MHz, $\Omega^{(0)}_{13}$=  45 MHz and $\Omega^{(0)}_{34}$=  16 MHz, shown to match experimental feature. The black curve also for stationary atoms with $\Omega^{(0)}_{12}$=  18 MHz, $\Omega^{(0)}_{13}$=  10 MHz and $\Omega^{(0)}_{34}$=  0.2 MHz.}
	\label{High_Rabi_exp}
\end{figure}

The transparency window splits when we increase the power of the 780 nm probe and blue control lasers as shown in Fig. \ref{High_Rabi_exp}a due to Autler-Townes splitting \cite{OSI_20}. 
With high power of 780 nm probe we see splitting of the main peak (near 0 MHz) and another small peak at $\sim$154 MHz as shown in Fig. \ref{High_Rabi_exp}a. The peak at $\sim$154 MHz is due to moving atoms with velocity $\sim$112 m/s along the direction of propagation of the 780~nm probe and blue control lasers. The Doppler shift for this velocity will bring the blue control laser in the resonance to $5S_{1/2} F=2$ $\rightarrow$ $6P_{1/2} F=1$ transition. The frequency of IR control laser for this velocity will be up-shifted by $\sim$ 111 MHz and we need to increase the frequency by another $\sim$154 MHz to bring the IR laser in the resonance with $6P_{1/2}F=1$ $\rightarrow$ ${69D_{3/2}}$ transition.

For this velocity group, the 780 nm probe laser will be down-shifted by $\sim$ 143 MHz and will be addressing $5S_{1/2} F=2$ $\rightarrow$ $5P_{3/2} F=3$ and $5S_{1/2} F=2$ $\rightarrow$ $5P_{3/2} F=2$ off resonantly due to power broadening effect. 

The splitting of the main peak is asymmetric because the 780 nm probe and blue control lasers are also driving other closely spaced hyperfine levels off-resonantly, causing a lightshift of  $5S_{1/2} F=2$ resulting in the detuning for the transitions. These effects can be incorporated by including the nearby hyperfine levels, ${5S_{1/2} F=2}$ and ${6P_{1/2} F=1}$  with following modified Hamiltonian.    

\begin{align}
	\label{eq_hyperfine}
	H&=\hbar[-\delta_{12}\ket{2}\bra{2}-\delta_{13}\ket{3}\bra{3}-(\delta_{13}+\delta_{34})\ket{4}\bra{4}\\ \nonumber
&-(\delta_{12}+HF1)\ket{5}\bra{5}-(\delta_{13}+HF2)\ket{6}\bra{6}\\\nonumber
	&+ \frac{\Omega_{12}}{2}\ket{1}\bra{2}+\frac{\Omega_{13}}{2}\ket{1}\bra{3}+\frac{\Omega_{34}}{2}\ket{3}\bra{4}\\\nonumber
&\frac{\Omega_{15}}{2}\ket{1}\bra{5}+\frac{\Omega_{16}}{2}\ket{1}\bra{6}+\frac{\Omega_{46}}{2}\ket{4}\bra{6}+h.c.]
\end{align}  

This Hamiltonian involves 36 coupled differential equations for the density matrix elements through Liouville- Von Neumann equation. We follow a similar numerical procedure as described for the four-level system to find out $\rho^{\textrm{ave}}_{21}$. We plot $\rho^{\textrm{ave}}_{21}$ vs IR laser detuning in Fig. \ref{High_Rabi_exp}b with the value of $\Omega^{(0)}_{12}$, $\Omega^{(0)}_{13}$, and $\Omega^{(0)}_{34}$ to match the spectral feature shown in Fig. \ref{High_Rabi_exp}a. We also plot in Fig. \ref{High_Rabi_exp}b the case for the stationary atom to see the ideal case.

\begin{figure}[ht!]
	\centering
   \includegraphics[width=0.45\linewidth]{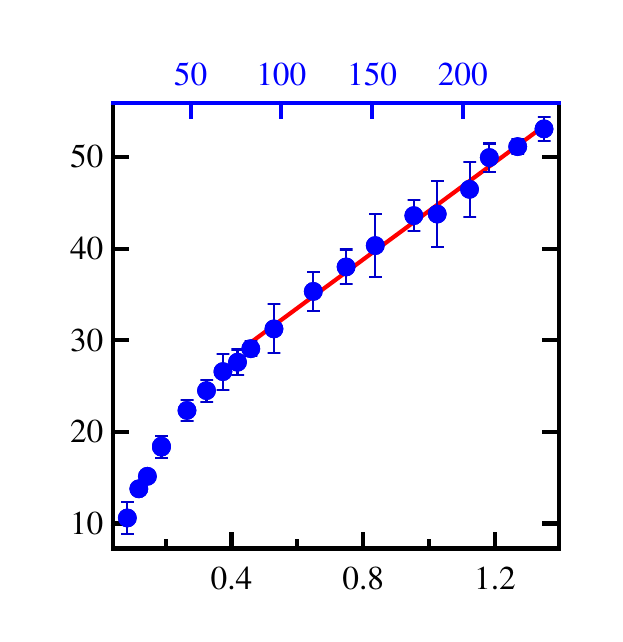}
    	\begin{picture}(0,0)
		\put(-120,100){(a)}	
		\put(-120,40){\rotatebox{90}{\centering $\Delta_{s}$(MHz)}}
		\put(-80,-5){$\sqrt{P_{780}}$$(\sqrt{mW})$}
        \put(-70,115){\textcolor{blue}{$\Omega^{(0)}_{12}$ (MHz)}}
	\end{picture}
     \includegraphics[width=0.45\linewidth]{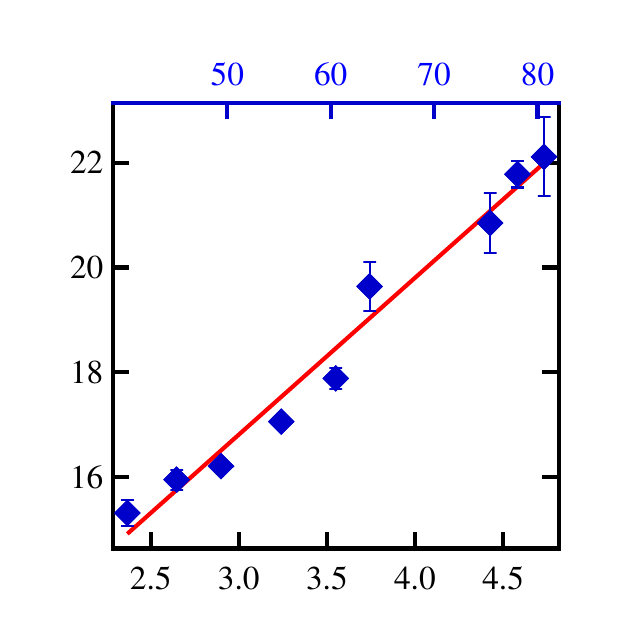}
   	\put(-110,40){\rotatebox{90}{\centering $\Delta_{s}$(MHz)}}
     \put(-80,-5){$\sqrt{P_{420}}$$(\sqrt{mW})$}
     \put(-70,115){\textcolor{blue}{$\Omega^{(0)}_{13}$ (MHz)}}
	  \put(-110,100){(b)}
	\caption{ \justifying The separation between doublet peaks measured under various \textbf{(a)} 780 nm probe power with blue control and IR control power of 7 mW ( $\Omega^{(0)}_{13}$=45 MHz) and 40 mW ($\Omega^{(0)}_{34}$=16 MHz) respectively, and \textbf{(b)} blue control power with 780 nm probe an IR control power of 20 $\mu$W ($\Omega^{(0)}_{12}$=25 MHz) and 40 mW ($\Omega^{(0)}_{34}$=16 MHz) respectively   }
	\label{fig:delta_420}
\end{figure}
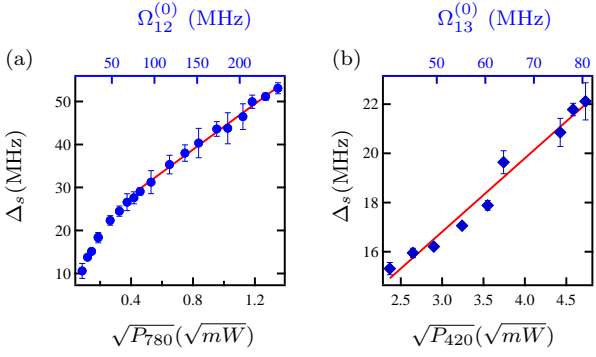

The absorption profile can be further understood by the dressed state picture. In the regime of Autler-Townes splitting, i.e., with high power of 780 nm probe and blue control lasers, there will be three transparency peaks corresponding to three dressed states, $\ket{0}=\frac{\Omega_{12}}{\sqrt{|\Omega_{12}|^2+|\Omega_{13}|^2}}\ket{3}-\frac{\Omega_{13}}{\sqrt{|\Omega_{12}|^2+|\Omega_{13}|^2}}\ket{2}$, $\ket{+}=\frac{1}{\sqrt{2}}\frac{\Omega_{12}}{\sqrt{|\Omega_{12}|^2+\frac{1}{2}|\Omega_{13}|^2}}\ket{3}+\frac{1}{\sqrt{2}}\frac{\Omega_{13}}{\sqrt{|\Omega_{12}|^2+|\Omega_{13}|^2}}\ket{2}+\frac{1}{\sqrt{2}}\ket{1}$; $\ket{-}=\frac{1}{\sqrt{2}}\frac{\Omega_{12}}{\sqrt{|\Omega_{12}|^2+\frac{1}{2}|\Omega_{13}|^2}}\ket{3}+\frac{1}{\sqrt{2}}\frac{\Omega_{13}}{\sqrt{|\Omega_{12}|^2+|\Omega_{13}|^2}}\ket{2}-\frac{1}{\sqrt{2}}\ket{1}$, with energy level $0$, $\frac{\sqrt{|\Omega_{12}|^2+|\Omega_{13}|^2}}{2}$ and $-\frac{\sqrt{|\Omega_{12}|^2+|\Omega_{13}|^2}}{2}$ respectively. The three peaks are visible in Fig. \ref{High_Rabi_exp}b for a stationary atom (black curve) with a certain combination of Rabi frequencies, but with averaging the central peak disappears. The separation between the two extreme transparency peaks will be $\sqrt{|\Omega_{12}|^2+|\Omega_{13}|^2}$. As the Rabi frequency happens to be proportional to the square root of power, we plot the separation between the doublet vs square root of power of the 780 nm probe and blue control laser in Fig. \ref{fig:delta_420}a and \ref{fig:delta_420}b respectively. When the Rabi frequency of one of the lasers is much higher than other laser it follows the linear behavior as shown by the red line in Fig. \ref{fig:delta_420}a and \ref{fig:delta_420}b.  
 
\begin{figure}[ht!]
	\centering
    \includegraphics[width=0.85\linewidth]{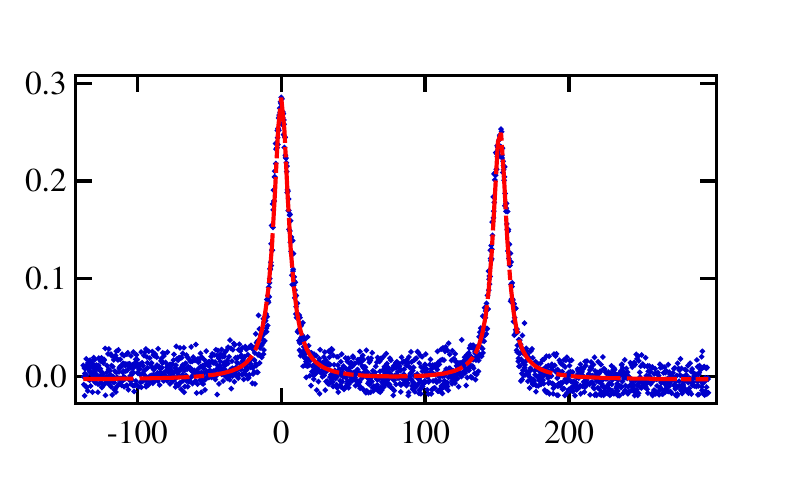}
	\begin{picture}(0,0)
		\put(-230,25){\rotatebox{90}{\shortstack{Probe transmission \\ (arb.units)}}}
		\put(-165,-1){Coupling detuning (MHz)}
	\end{picture}
	\caption{ \justifying The observed probe transmission signal (after averaging of eight traces) for the excitation $5S_{1/2}$ ($F=2$) $ \rightarrow$ $6P_{1/2}$  $\rightarrow$ $69D_{3/2}$ with probe and coupling power of 60 $\mu$W and 65 mW respectively. The right and left transmission peak corresponds to the hyperfine levels of  $6P_{1/2}$ state, $F = 2$ and $F = 1$, respectively. The blue dot (red dashed line) is the experimental data (Lorentzian fit). }
	\label{spectrum_inv_ladder}
\end{figure}
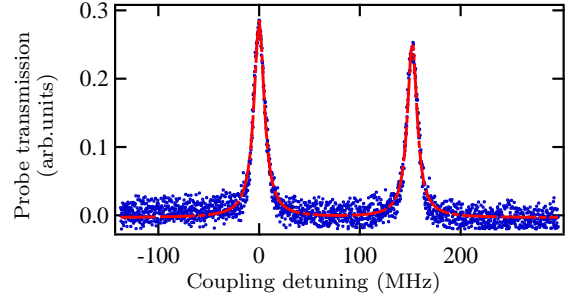

Now, we discuss the inverted $\Xi$-system where the probe at 421 nm is driving the ${5S_{1/2} F=2}\rightarrow {6P_{1/2} F=2}$ transition and IR control laser is scanning around ${6P_{1/2} F=1, 2}\rightarrow {69D_{3/2}}$.  
The transmission peak on left side in Fig. \ref{spectrum_inv_ladder} corresponds zero velocity group of atom and the transition ${6P_{1/2} F=2} \rightarrow{69D_{3/2}}$. The right one corresponds to $v=112$ m/s along propagation direction of blue probe laser and the transition ${6P_{1/2} F=1}\rightarrow {69D_{3/2}}$ transition. The signal-to-noise ratio for spectrum is 36 after eight averaging, $<$1 without averaging and 4 for the error signal without averaging.     

\begin{figure}[ht!]
	 \centering
	\includegraphics[width=0.5\linewidth]{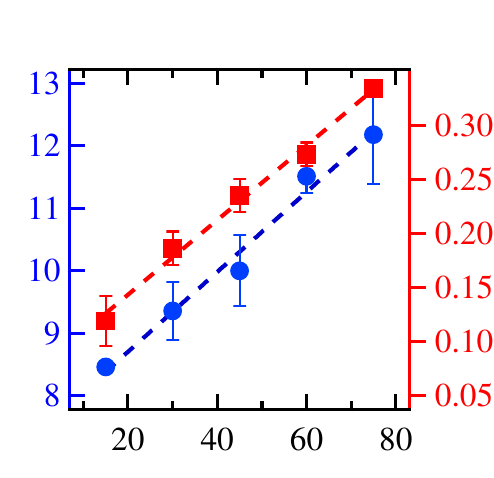}
	\begin{picture}(0,0)
		\put(-140,30){\rotatebox{90}{Linewidth (MHz)}}
		\put(-110,-5){Coupling power (mW)}
        \put(10,40){\rotatebox{90}{H (arb.units)}}
	\end{picture}
	\caption{  \label{fig:Pvslh} \justifying  The experimental linewidth (blue circle) and the transmission peak height (H) (Red square) of the transmission peak attributed to the transition $5S_{1/2}$ ($F = 2$) $\rightarrow$ $6P_{1/2} (F=2)$  $\rightarrow$ $69D_{3/2}$ as a function of the power of the coupling beam with probe power of 60 $\mu$W. Dashed lines blue and red are the line fitting for the linewidth and peak height.}
\end{figure}
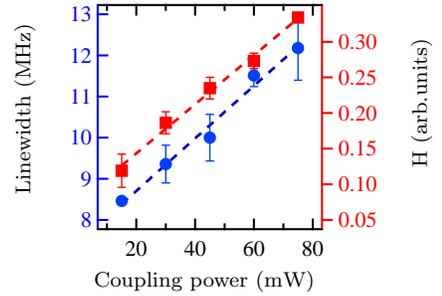

 To extract the linewidth and peak height, the spectrum is fitted with the sum of two Lorentzian functions. The linewidth and the transmission peak height (H) of the peak corresponding to the transition from ${6P_{1/2} F=2}\rightarrow {69D_{3/2}}$ with various coupling powers are shown in Fig \ref{fig:Pvslh}. The linewidth increases with increasing coupling power due to the power broadening mechanism which fits well to the line. The minimum observed linewidth is around 8 MHz.

In conclusion, we have experimentally investigated the EIT spectrum using the Rydberg state in thermal Rb atoms with  V+inverted $\Xi$ system using ${5S_{1/2}}$ $\rightarrow$ ${5P_{3/2}}$ and ${5S_{1/2}}$ $\rightarrow$ ${6P_{1/2}}$  $\rightarrow$ ${69D_{3/2}}$.  For V+inverted $\Xi$ system the probe laser at 780 nm gives good EIT spectrum with scan of IR Rydberg laser even without heating of the Rb cell and does not require blue sensitive photodetector to observe the EIT spectrum. The linewidth of the EIT spectrum is around 9 MHz. In comparison to the EIT spectrum in inverted $\Xi$ (which requires heating of the Rb cell and blue-sensitive photodetector), the V+inverted $\Xi$ system has a better signal-to-noise ratio with similar linewidth. The study will be very useful to stabilize (long term) the IR Rydberg laser for quantum computation and simulation and other proposed applications based upon the V+ $\Xi$ systems.     

\textit{Acknowledgement}-R.C.D. acknowledges the Ministry
of Education, Government of India, for the Prime Minister’s Research Fellowship (PMRF), and K.P. acknowledges funding from DST through Grant No. DST/QTC/NQM/QC/2024/1 (G).
\bibliography{Eit_rb}

\end{document}